\documentclass{Interspeech}
\interspeechcameraready 

\title{Hearing from Silence: Reasoning Audio Descriptions from \\Silent Videos via Vision-Language Model\thanks{$^\dagger$ Corresponding author}}

\author[affiliation={1,2,3}]{Yong}{Ren}
\author[affiliation={3},correspondence={\dagger}]{Chenxing}{Li}
\author[affiliation={1}]{Le}{Xu}
\author[affiliation={1,2}]{Hao}{Gu}
\author[affiliation={1}]{Duzhen}{Zhang}
\author[affiliation={1}]{Yujie}{Chen}
\author[affiliation={3}]{Manjie}{Xu}
\author[affiliation={1}]{\\Ruibo}{Fu}
\author[affiliation={3}]{Shan}{Yang}
\author[affiliation={4},correspondence={\dagger}]{Dong}{Yu}

\affiliation{Institute of Automation}{Chinese Academy of Sciences}{Beijing, China\vskip1pt}
\affiliation{School of Artificial Intelligence}{University of Chinese Academy of Sciences}{Beijing, China\vskip1pt}
\affiliation{}{Tencent AI Lab}{Beijing, China\ \ }
\affiliation{}{Tencent AI Lab}{Seattle, USA}
\email{lichenxing007@gmail.com, dongyu@ieee.org}


\usepackage[table]{xcolor}
\usepackage{comment}
\usepackage{multirow}
\usepackage{arydshln}
\usepackage{graphicx}
\usepackage{subfigure}
\usepackage{algorithm}
\usepackage{algorithmic}
\usepackage{cite}
\usepackage{makecell}

\definecolor{LightCyan1}{rgb}{0.88,1,1}
\definecolor{deepred}{rgb}{0.698,0.133,0.133}
\definecolor{shadecolor}{rgb}{0.92,0.92,0.92}
\definecolor{deepgreen}{rgb}{0.23,0.49,0.14}

\begin{document}

\maketitle

\begin{abstract}
Humans can intuitively infer sounds from silent videos, but whether multimodal large language models can perform modal-mismatch reasoning without accessing target modalities remains relatively unexplored. Current text-assisted-video-to-audio (VT2A) methods excel in video foley tasks but struggle to acquire audio descriptions during inference. We introduce the task of Reasoning Audio Descriptions from Silent Videos (SVAD) to address this challenge and investigate vision-language models' (VLMs) capabilities on this task. To further enhance the VLMs' reasoning capacity for the SVAD task, we construct a CoT-AudioCaps dataset and propose a Chain-of-Thought-based supervised fine-tuning strategy. Experiments on SVAD and subsequent VT2A tasks demonstrate our method's effectiveness in two key aspects: significantly improving VLMs' modal-mismatch reasoning for SVAD and effectively addressing the challenge of acquiring audio descriptions during VT2A inference.

\end{abstract}

\section{Introduction}

\begin{figure}[t]
\centering
  \centerline{\includegraphics[width=1.0\linewidth]{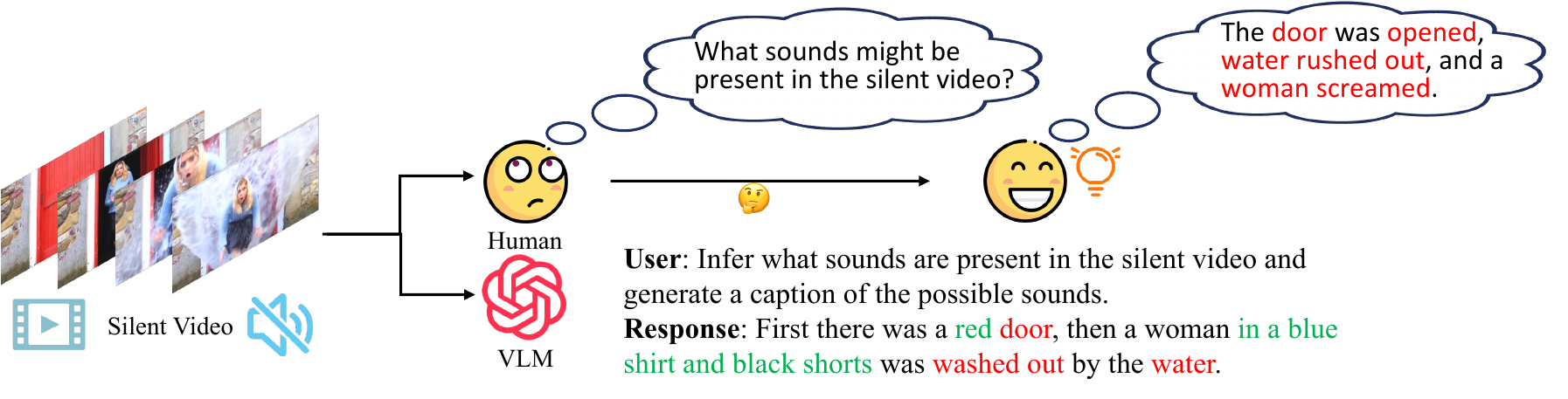}}
  \caption{
  Good sound descriptions from humans. vs. auditory-irrelevant hallucination from VLMs when reason audio descriptions from silent videos.
  }
  \vspace{-5pt}
  \label{fig:intro1}
\end{figure}

Human cognition inherently integrates multimodal information, allowing us to infer auditory experiences from purely visual stimuli like silent videos as shown in Figure \ref{fig:intro1}. This remarkable ability stems from our brain's capacity to associate visual patterns with corresponding sounds through learned experiences and cognitive reasoning \cite{han2022multisensory}. Although recent advancements in multimodal large language models (MLLMs) have demonstrated impressive capabilities in multimodal understanding and reasoning \cite{zhang2024mm,li2023videochat,lin2023video,ataallah2024minigpt4}, their ability of modal-mismatch reasoning in the absence of target modalities remains largely unexplored.

\begin{figure}[t]
\centering
  \centerline{\includegraphics[width=1.0\linewidth]{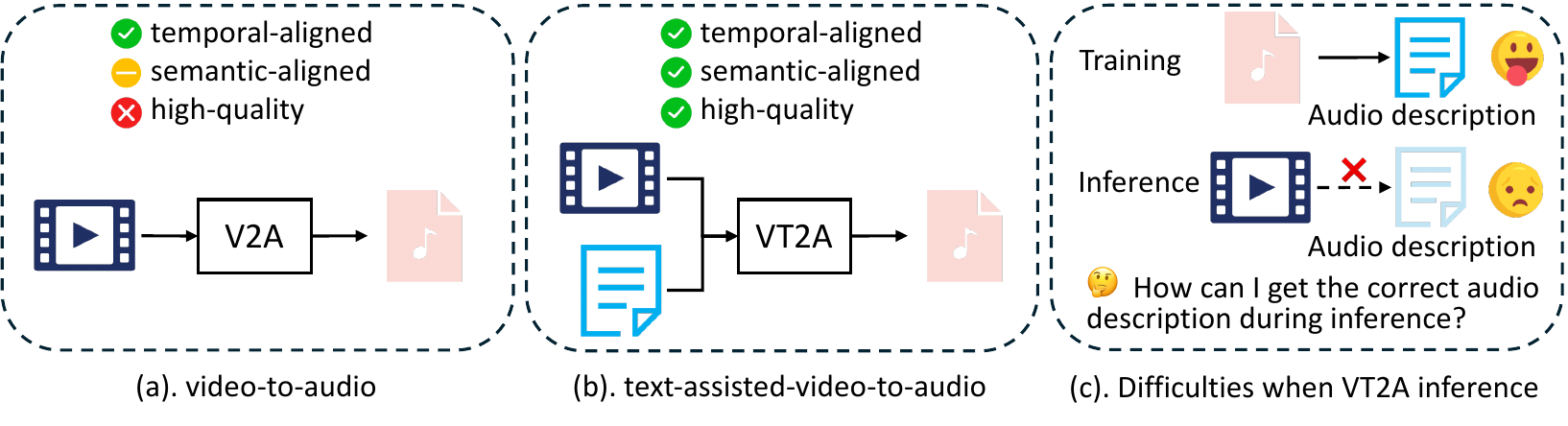}}
  \caption{
  Two primary technical paradigms of video foley and challenges faced by VT2A.
  }
  \vspace{-15pt}
  \label{fig:intro2}
\end{figure}

How to reason unseen-modality-related information is not only of significant exploratory value for advancing MLLMs towards more human-like capabilities but also holds important implications in practical applications such as video foley. As illustrated in Figure \ref{fig:intro2} (a) and (b), current video foley approaches primarily follow two technical paradigms: Video-to-Audio (V2A) \cite{xu2024video,wang2024frieren,wang2024v2a,mei2024foleygen}, which generate audio solely from visual input, and text-assisted-video-to-audio (VT2A) \cite{ren2024sta,mo2024text,zhang2024foleycrafter,jeong2024read,gramaccioni2024stable}, which uses textual descriptions as additional guidance. Although VT2A methods outperform V2A regarding semantic consistency and audio quality, they encounter a significant challenge during inference, as shown in Figure \ref{fig:intro2} (c). Typically, VT2A models only receive silent videos without the corresponding textual descriptions of the target audio during inference, necessitating manual annotations by human experts. To address this challenge, we introduce the Reasoning Audio Descriptions from Silent Videos (SVAD) task. Unlike existing caption tasks such as audio caption \cite{mei2024wavcaps,labb2024conette}, video caption \cite{kavitha2024automatic,zhou2024streaming}, and audio-visual caption\cite{kim2024avcap,rho2025lavcap}, SVAD challenges on reasoning information related to a modality (audio) that does not match the input modality (visual).

\begin{figure*}[t]
\centering
  \centerline{\includegraphics[width=1.0\linewidth]{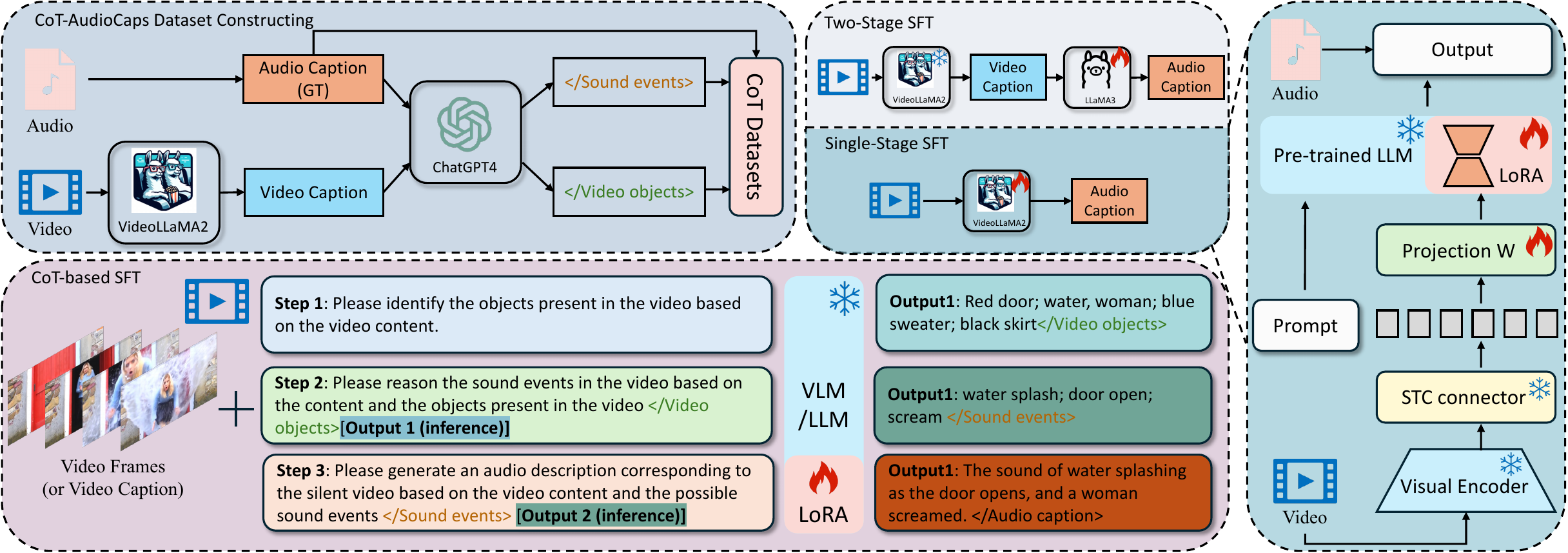}}
  \caption{
  Overview of our methods for SVAD task, including two SFT strategies, the SFT training for VLM by LoRA, the CoT-Audiocaps Dataset construction process, and the CoT-based SFT method for SVAD.
  }
  \label{fig:main}
  \vspace{-10pt}
\end{figure*}

AVCap\cite{kim2024avcap} can be adapted for the SVAD task when trained with only the video modality. 
DALI\cite{malard2025eye} can align the distributions of visual and auditory modalities through training, allowing the substitution of image encodings with aligned audio encodings for SVAD tasks.
However, their performance in the SVAD task is limited and insufficient to substitute for the audio captions required during VT2A inference.
Recent advancements in VLMs have demonstrated remarkable capabilities in video understanding and reasoning tasks\cite{cheng2024videollama,liu2024oryx,chen2024internvl}. Therefore, we attempt to use state-of-the-art (SOTA) VLMs to tackle the SVAD challenge. This also serves as an effective evaluation of VLMs' modal-mismatch reasoning abilities.

Our evaluation across multiple SOTA VLMs reveals that pre-trained models, even the best-performing VideoLLaMA2 \cite{cheng2024videollama}, show suboptimal results on the SVAD task, highlighting the need for specialized enhancement strategies. To address this limitation, we employ supervised fine-tuning (SFT) by Low-Rank Adaptation (LoRA) \cite{hulora}, a prevailing technique for improving LLMs' reasoning capabilities through task-specific adaptation. We design two distinct SFT approaches: (1) a two-stage strategy where the pre-trained VLM first generates detailed video descriptions, followed by fine-tuning LLM to derive audio descriptions from these visual narratives; and (2) a single-stage strategy that directly fine-tunes VLM using audio descriptions as ground truth (GT). SFT has shown a significant improvement in SVAD tasks, with the single-stage strategy performing better. Chain-of-thought (CoT) is a specialized tool designed for the task of multi-step reasoning and decision-making.\cite{wei2022chain} To further enhance the reasoning capabilities in SVAD tasks, we propose a Chain-of-Thought-based Supervised Fine-Tuning (CoT-SFT) strategy and construct the CoT-AudioCaps dataset for it, which provides explicit reasoning chains connecting visual scenes to their corresponding audio descriptions. This approach enables VLMs to systematically decompose the SVAD task into three coherent stages: visual object understanding, sound event reasoning, and audio description prediction. The CoT-SFT strategy showed superior performance in SVAD tasks. Finally, we validated the effectiveness of our method on two SOTA VT2A methods. Our contributions can be summarized as follows:
\begin{itemize}
\item We propose the SVAD task designed to address the problem of missing audio descriptions during VT2A inference.
\item We explore VLMs' modal-mismatch reason capabilities by the SVAD task.
\item We propose a CoT-based SFT strategy for the SVAD task and construct the CoT-AudioCaps dataset, significantly enhancing VLMs' modal-mismatch reasoning capabilities.
\item Experimental results demonstrate that our method effectively improves performance in the SVAD task and addresses the audio description acquisition challenge in VT2A inference.
\end{itemize}

\section{Methods}
\label{methods}

\subsection{SFT for SVAD}

Given a silent video $V={I}_{t=1}^T$ with $T$ frames, the SVAD task aims to generate a corresponding audio description $C_{\text{audio}}$.
\begin{equation}
C_{\text{audio}} = \mathcal{F}(V),
\end{equation}
where $\mathcal{F}$ denotes the vision understanding and reasoning models like VLMs. 
Utilizing pre-trained VLMs for zero-shot inference often results in suboptimal performance. 
Existing VLMs are typically pre-trained on multimodal alignment tasks, so they fail to address modal-mismatch reasoning when the target modality (audio) is absent. 
As shown in Figure \ref{fig:intro1}, pre-trained VLMs tend to generate auditory-irrelevant information such as color, shape, and size, while overlooking implicit sound events, such as a woman screaming.
To address this, we utilize pairs of audio descriptions and video to perform SFT, and design two strategies: 

\textbf{Two-Stage SFT:} Decouples visual perception (VLM zero-shot inference) and audio reasoning (LLM SFT) :
\begin{align}
C_{\text{video}} &= VLM(V),  \\
C_{\text{audio}} &= LLM(C_{\text{video}}; \theta_{\text{LoRA}}) .
\end{align}

\textbf{Single-Stage SFT:} Jointly optimizes perception and reasoning through VLM SFT:
\begin{equation}
C_{\text{audio}} = VLM(V; \theta_{\text{LoRA}}),
\end{equation}
where $\theta_{\text{LoRA}}$ denotes weights of LoRA Adapters.

\begin{figure*}[t]
\centering
  \centerline{\includegraphics[width=0.95\linewidth]{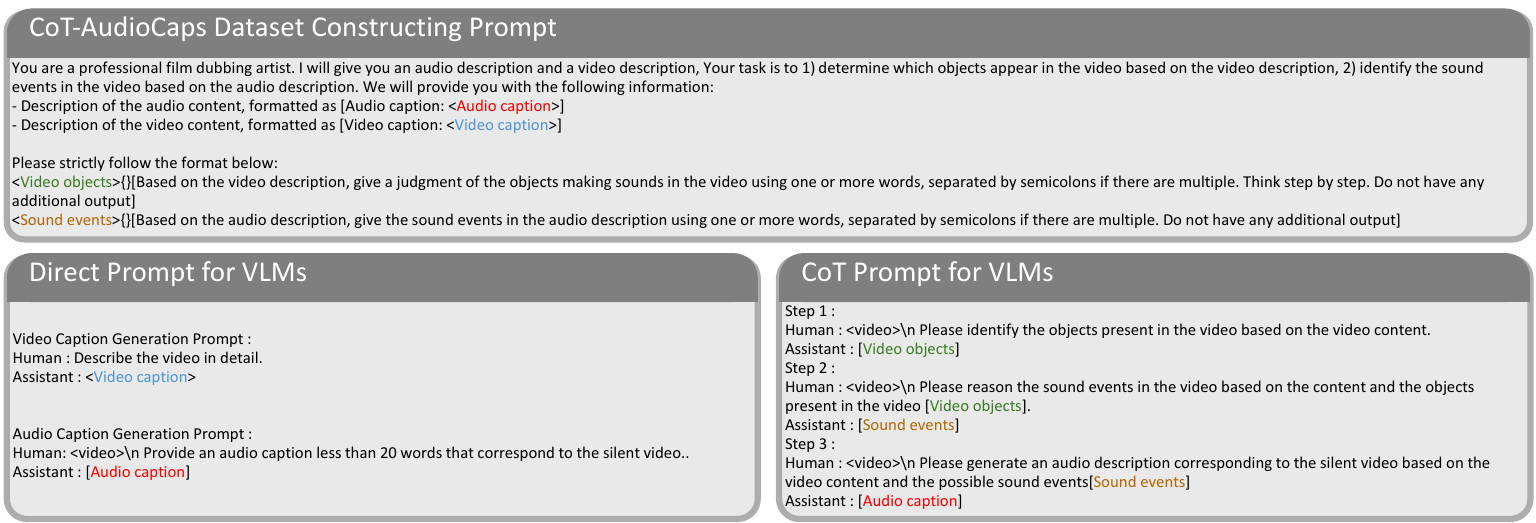}}
  \caption{
  The templates for constructing the CoT-Audiocaps Dataset, the direct prompt template for video and audio caption from video for VLMs, and the CoT prompt template for VLMs (For LLMs, replace the video with the video caption).
  }
  \label{fig:prompt}
  \vspace{-8pt}
\end{figure*}

\subsection{CoT-AudioCaps: Dataset Construction}
\label{cot-audiocaps}

We construct the CoT-AudioCaps dataset through VideoLLAMA2 and GPT-4 from the AudioCaps dataset\cite{kim2019audiocaps}.

\begin{algorithm}[H]
\caption{CoT-AudioCaps Dataset Construction}
\label{alg}
\begin{algorithmic}[1]
\renewcommand{\algorithmicrequire}{\textbf{INPUT}}
\renewcommand{\algorithmicensure}{\textbf{OUTPUT}}
\REQUIRE Audioset dataset $\mathcal{D}=\{(V^k,C_{\text{audio}}^k)\}_{k=1}^{|D|}$, VLM $\mathcal{VLM}$, LLM $\mathcal{LLM}$, video caption prompt template $\mathcal{P}_{vc}^{user}$, CoT information acquisition prompt template $\mathcal{P}_{reason}^{user}$
\ENSURE Visual to Video Object dataset: $\mathcal{D}_{v2o}$, Video Object to Sound Event dataset: $\mathcal{D}_{o2e}$, Sound Event to Audio Caption dataset $\mathcal{D}_{e2c}$
\FOR{each ($V$, $C_{\text{audio}}$) $\in \mathcal{D}$}
    \STATE $C_{\text{video}} = \mathcal{VLM}(\mathcal{P}_{vc}^{user}(V))$
    \STATE $<V_{\text{object}}, S_{\text{event}}>=\mathcal{LLM}(\mathcal{P}_{reason}^{user}(C_{\text{video}}, C_{audio}))$
    \STATE $\mathcal{D}_{v2o} +=\{(V/C_{\text{video}}),(V_{\text{object}})\}$
    \STATE $\mathcal{D}_{o2e} +=\{(V/C_{\text{video}},V_{\text{object}}),(S_{\text{event}})\}$
    \STATE $\mathcal{D}_{e2c} +=\{(V/C_{\text{video}},S_{\text{event}}),(C_{\text{audio}})\}$
\ENDFOR
\end{algorithmic}
\end{algorithm}

\begin{table*}[ht]
    \caption{Evaluation of several SOTA pre-trained VLMs in SVAD. The \textcolor{deepred}{\textbf{red}} highlights the highest performance, and the \textcolor{blue}{\textbf{blue}} indicates the second-highest performance.}
    \label{tab:results0}
    \centering
    \setlength{\tabcolsep}{0.9mm}{
    \scalebox{0.92}{
    \begin{tabular}{lccccccccc} 
        \toprule
        \multirow{2}{*}{\textbf{VLM}} & \textbf{Text-Audio} & \multicolumn{8}{c}{\textbf{Text-Text}} \\ 
        \cline{3-10} 
            & CLAP\textcolor{red}{$\uparrow$}  & BLEU\_1\textcolor{red}{$\uparrow$} & BLEU\_2\textcolor{red}{$\uparrow$} & BLEU\_3\textcolor{red}{$\uparrow$} & BLEU\_4\textcolor{red}{$\uparrow$} & METEOR\textcolor{red}{$\uparrow$} & ROUGE\_L\textcolor{red}{$\uparrow$} & CIDEr\textcolor{red}{$\uparrow$} & SPICE\textcolor{red}{$\uparrow$} \\ \midrule
        GT                  & 0.591 & 1.000 & 1.000 & 1.000 & 1.000 & 1.000 & 1.000 & 2.658 & 0.429 \\
        \cdashline{1-10}
        GPT-4o\cite{hurst2024gpt}               & 0.244 & 0.249 & 0.086 & 0.034 & 0.014 & 0.083 & 0.197 & 0.088 & 0.047 \\
        Oryx \cite{liu2024oryx} (7B)    
        & 0.193                                 & \textcolor{blue}{\textbf{0.302}}                              
        & \textcolor{blue}{\textbf{0.131}}      & \textcolor{blue}{\textbf{0.055}}                              
        & \textcolor{blue}{\textbf{0.023}}      & 0.102                                                         
        & \textcolor{blue}{\textbf{0.258}} & 0.118 & 0.054 \\
        InternVL2.5 \cite{chen2024internvl} (8B)    
        & \textcolor{deepred}{\textbf{0.254}}   & 0.293                                                         
        & 0.126                                 & \textcolor{blue}{\textbf{0.055}}                              
        & 0.021                                 & \textcolor{blue}{\textbf{0.109}}                              
        & 0.256                                 & \textcolor{blue}{\textbf{0.119}}                              
        & \textcolor{blue}{\textbf{0.060}} \\
        \rowcolor{LightCyan1}
        VideoLLaMA2 \cite{cheng2024videollama} (7B)    
        & \textcolor{blue}{\textbf{0.252}}      & \textcolor{deepred}{\textbf{0.387}}                           
        & \textcolor{deepred}{\textbf{0.194}}   & \textcolor{deepred}{\textbf{0.092}}                           
        & \textcolor{deepred}{\textbf{0.041}}   & \textcolor{deepred}{\textbf{0.128}}                           
        & \textcolor{deepred}{\textbf{0.302}}   & \textcolor{deepred}{\textbf{0.182}}                           
        & \textcolor{deepred}{\textbf{0.073}} \\
        \bottomrule 
    \end{tabular}
    }
    }
    \vspace{-10pt}
\end{table*}

As detailed in Algorithm \ref{alg} and Figure \ref{fig:main}, the pipeline operates as follows: For each video-audio\_caption pair ($V,C_{\text{audio}}$), we first get video captions $C_{\text{video}}$ by the pre-trained VLM (VideoLLaMA2); then we use The LLM (GPT-4) parses $C_{\text{video}}$ and $C_{\text{audio}}$ to extract structured reasoning components including video objects $V_{\text{object}}$ and sound events $S_{\text{event}}$; finally we use V (for Single-Stage)/$C_{\text{video}}$ (for Two-Stage), $V_{\text{object}}$, $S_{\text{event}}$ and $C_{\text{audio}}$ to construct the CoT-AudioCaps Dataset for CoT-based SFT. The details of the prompts are shown in Figure \ref{fig:prompt}.

\subsection{CoT-based SFT: SFT strategy}
We propose a CoT-based SFT method designed for the SVAD task, which enhances the model's reasoning capabilities and interpretability by decomposing SVAD into three subtasks. Utilizing the CoT-AudioCaps dataset obtained in Section \ref{cot-audiocaps} and the prompt shown in Figure \ref{fig:prompt}, we perform SFT as follows:
\begin{align}
V_{\text{object}} &= VLM(V; \theta_{\text{LoRA}}), \\
S_{\text{event}} &= VLM(V, V_{\text{object}}; \theta_{\text{LoRA}}),  \\
C_{\text{audio}} &= VLM(V, S_{\text{event}}; \theta_{\text{LoRA}}).
\end{align}
Subtask 1 involves reasoning video objects $V_{\text{object}}$ from video $V$, Subtask 2 involves reasoning sound events $S_{\text{event}}$ from $V$ and $V_{\text{object}}$, and Subtask 3 involves reasoning audio descriptions $C_{\text{audio}}$ from $V$ and $S_{\text{event}}$. Taking Single-stage SFT as an example, during training, the VLM is fine-tuned using the CoT-AudioCaps dataset $\mathcal{D}=\{\mathcal{D}_{v2o}, \mathcal{D}_{o2e}, \mathcal{D}_{e2c}\}$ by LoRA. During inference, the outputs $V_{\text{object}}$ and $S_{\text{event}}$ for Subtask 2 and 3 respectively use the output of the previous subtask. For Two-Stage SFT, simply replace $V$ with $C_{\text{video}}$ and VLM with LLM.

\section{Experiments}
\label{sec:exp}

In this section, we conduct detailed experiments to evaluate the performance of VLMs in SVAD, and the effectiveness of the proposed method on SVAD and VT2A tasks.
Our experiments seek to answer the following research questions (RQs):
\begin{itemize}
    \item \textbf{RQ1}: How do different pre-training VLMs perform in modal-mismatch reasoning for the SVAD task?
    \item \textbf{RQ2}: Is SFT effective for solving the SVAD task? Which of the two SFT strategies is more effective? Can our proposed CoT-based SFT further improve performance?
    \item \textbf{RQ3}: Can better audio descriptions obtained from silent videos reduce performance loss during VT2A inference?
\end{itemize} 

\begin{table*}[t]
    \caption{Results of the VideoLLaVA2 (VL2) and LLaMA3(LM3) backbone in SVAD. 
    $*$ indicates citing from the original paper.
    }
    \label{tab:results1}
    \centering
    \setlength{\tabcolsep}{0.9mm}{
    \scalebox{0.9}{
    \begin{tabular}{llccccccccc} 
        \toprule
        \multirow{2}{*}{\textbf{Strategy}} & \multirow{2}{*}{\textbf{Method}} & \textbf{Text-Audio} & \multicolumn{8}{c}{\textbf{Text-Text}} \\ 
        \cline{4-11} 
            & & CLAP\textcolor{red}{$\uparrow$} & BLEU\_1\textcolor{red}{$\uparrow$} & BLEU\_2\textcolor{red}{$\uparrow$} & BLEU\_3\textcolor{red}{$\uparrow$} & BLEU\_4\textcolor{red}{$\uparrow$} & METEOR\textcolor{red}{$\uparrow$} & ROUGE\_L\textcolor{red}{$\uparrow$} & CIDEr\textcolor{red}{$\uparrow$} & SPICE\textcolor{red}{$\uparrow$} \\ \midrule
        & GT                    & 0.591 & 1.000 & 1.000 & 1.000 & 1.000 & 1.000 & 1.000 & 2.658 & 0.429 \\
        \cdashline{1-11}
        & AVCap-V* \cite{kim2024avcap}
                                & -     & -     & -     & 0.247 & 0.158 & 0.153 & 0.391 & 0.441 & 0.107 \\
        & $\text{DALI}_{\text{OT}}^{\text{Att}}$* \cite{malard2025eye}       & -     & -     & -     & -     & 0.082 & 0.128 & 0.311 & 0.244 & 0.074 \\
        \cdashline{1-11}
        \multirow{3}{*}{Two-Stage}
        & VL2 + LM3             & 0.039 & 0.199 & 0.079 & 0.025 & 0.007 & 0.072 & 0.164 & 0.013 & 0.017 \\
        & VL2 + LM3-SFT         & 0.348 & 0.547 & 0.356 & 0.219 & 0.122 & 0.171 & 0.376 & 0.417 & 0.114 \\
        \rowcolor{shadecolor}
        & VL2 + LM3-CoT-SFT     & 0.373 & 0.554 & 0.363 & 0.228 & 0.134 & 0.178 & 0.381 & 0.424 & 0.115 \\
        \cdashline{1-11}
        \multirow{3}{*}{Single-Stage}
        & VL2                   & 0.252 & 0.387 & 0.194 & 0.092 & 0.041 & 0.128 & 0.302 & 0.182 & 0.073 \\
        & VL2-SFT               
        & \textcolor{blue}{\textbf{0.404}}      & \textcolor{deepred}{\textbf{0.633}}   
        & \textcolor{blue}{\textbf{0.438}}      & \textcolor{blue}{\textbf{0.286}}      
        & \textcolor{blue}{\textbf{0.172}}      & \textcolor{blue}{\textbf{0.195}}      
        & \textcolor{blue}{\textbf{0.436}}      & \textcolor{blue}{\textbf{0.550}}    
        & \textcolor{deepred}{\textbf{0.141}} \\
        \rowcolor{LightCyan1}
        & VL2-CoT-SFT           
        & \textcolor{deepred}{\textbf{0.424}}   & \textcolor{blue}{\textbf{0.618}}      
        & \textcolor{deepred}{\textbf{0.442}}   & \textcolor{deepred}{\textbf{0.298}}   
        & \textcolor{deepred}{\textbf{0.185}}   & \textcolor{deepred}{\textbf{0.196}}   
        & \textcolor{deepred}{\textbf{0.439}}   & \textcolor{deepred}{\textbf{0.578}} 
        & \textcolor{blue}{\textbf{0.130}}    \\
        \bottomrule 
    \end{tabular}
    }
    }
    \vspace{-10pt}
\end{table*}

\begin{table}[ht]
    \caption{Results on VT2A task using different audio descriptions (AD) as text prompts during inference.}
    \label{tab:results2}
    \centering
    \setlength{\tabcolsep}{0.9mm}{
    \scalebox{0.82}{
    \begin{tabular}{llccccc} 
        \toprule
        Method & AD & FD \textcolor{deepgreen}{$\downarrow$} & FAD \textcolor{deepgreen}{$\downarrow$} & KL \textcolor{deepgreen}{$\downarrow$}& IS \textcolor{red}{$\uparrow$} & AV-Align \textcolor{red}{$\uparrow$} \\ 
        \midrule
        \multirow{4}{*}{\makecell[c]{STA-V2A \\ \cite{ren2024sta}}}
        & GT                & 21.99 &  3.56 &  4.18 &  7.87 & 0.244 \\
        \cdashline{2-7}
        & w/o AD            & 44.07 &  9.62 & 11.28 &  4.46 & 0.210 \\
        & VL2               & 29.41 &  5.98 &  6.95 &  7.55 & 0.232 \\
        \rowcolor{LightCyan1}
        & VL2-CoT-SFT
        & \textcolor{deepred}{\textbf{23.43}}   & \textcolor{deepred}{\textbf{2.80}} 
        & \textcolor{deepred}{\textbf{5.11}}    & \textcolor{deepred}{\textbf{7.67}}
        & \textcolor{deepred}{\textbf{0.243}} \\
        \cline{1-7}
        \multirow{4}{*}{{\makecell[c]{FoleyCraft \\ \cite{zhang2024foleycrafter}}}}
        & GT                & 14.57 &  2.51 &  3.16 & 13.88 & 0.232 \\
        \cdashline{2-7}
        & w/o AD            & 21.70 &  3.32 &  5.87 & 10.68 & 0.233 \\
        & VL2               & 21.61 &  3.27 &  5.63 & 12.88 & 0.234 \\
        \rowcolor{LightCyan1}
        & VL2-CoT-SFT       
        & \textcolor{deepred}{\textbf{21.07}} & \textcolor{deepred}{\textbf{2.94}} 
        & \textcolor{deepred}{\textbf{4.74}} & \textcolor{deepred}{\textbf{13.28}}  
        & \textcolor{deepred}{\textbf{0.243}} \\
        \bottomrule 
    \end{tabular}
    }
    }
    \vspace{-10pt}
\end{table}

\subsection{Experimental Settings}

\subsubsection{Datasets and Baselines}
We use the AudioCaps \cite{kim2019audiocaps} dataset, which contains 43,941 training instances, 447 validation instances, and 866 evaluating instances with videos and audio captions annotation. We adopt the ablation experimental results from AVCap \cite{kim2024avcap} that uses only video features (AVCap-V) and the results of best alignment method $\text{DALI}_{\text{OT}}^{\text{Att}}$ in \cite{malard2025eye} as our baselines. 

\subsubsection{Metrics}
We use CLAP\footnote{https://huggingface.co/lukewys/laion\_clap/blob/main/630k-best.pt} \cite{wu2023large} to compute the embedding similarity between text and audio.
For similarity between text and text, we use traditional captioning metrics focusing on token-level matching, including BLEU \cite{papineni2002bleu}, METEOR \cite{banerjee2005meteor}, ROGUEl \cite{lin2004rouge}, 
CIDEr \cite{vedantam2015cider}, and SPICE \cite{anderson2016spice}. For VT2A evaluation, we use Fr\'{e}chet distance distance (FD), Fr\'{e}chet Audio Distance (FAD), KL divergence (KL), Inception Score (IS), and AV-Align\cite{yariv2024diverse}.

\subsubsection{Implementation Details.}
We utilize VideoLLaMA2\footnote{https://github.com/DAMO-NLP-SG/VideoLLaMA2} \cite{cheng2024videollama} and LLaMA3\footnote{https://github.com/hiyouga/LLaMA-Factory} \cite{dubey2024llama} as our backbone, incorporating LoRA \cite{hulora} into them. During the VideoLLaMA2 SFT, both the vision encoder and the LLM remain frozen, with only the projector and LoRA components being trained. The LoRA rank $r$ is set to $128$, the $\alpha$ is set to $256$, and the learning rate $lr$ is set to $2e-5$.
We utilize STA-V2A\footnote{https://github.com/y-ren16/STAV2A} \cite{ren2024sta} and FoleyCraft\footnote{https://github.com/open-mmlab/FoleyCrafter} \cite{zhang2024foleycrafter} as VT2A model. All experiments are conducted on $4$ NVIDIA $40$GB A100 GPUs.

\subsection{Eval Pre-training VLMs in SVAD (RQ1)}
We evaluated the performance of various models on the SVAD task, employing GPT-4o and three SOTA pre-trained VLMs. 
The pre-trained VLMs include Oryx-1.5-7B\footnote{https://github.com/Oryx-mllm/Oryx}, InternVL2.5-8B\footnote{https://github.com/OpenGVLab/InternVL}, and VideoLLaMA2.1-7B-16F.
The prompts used for this evaluation are depicted in the "Direct Prompt" of Table \ref{tab:results0}.

The results presented in Table 1 indicate that \textbf{directly using these pre-trained VLMs to address the SVAD task generally results in poor performance}. This underperformance can be attributed to the task's dual requirements: understanding visual content and executing modal-mismatch reasoning, which collectively poses a significant challenge. Among the models evaluated, VideoLLaMA2 performed the best in the Text-Text similarity metric, and on the Text-Audio similarity metric CLAP, it was only 0.002 lower than InternVL. Therefore, \textbf{VideoLLaMA2 exhibited the best overall performance} and was selected as the VLM backbone for subsequent experiments.

\subsection{SFT and CoT-based SFT (RQ2)}
We explored Two-Stage and Single-Stage strategies for addressing the SVAD task. According to results shown in Table \ref{tab:results1}, \textbf{the performance under the One-Stage strategy consistently surpassed that of the Two-Stage strategy}. This is attributed to the fact that although the Two-Stage strategy leverages the reasoning capabilities of LLMs, the process of converting videos to video captions inherently results in some information loss. \textbf{After performing SFT with GT audio captions, there was a significant improvement in the model's performance on the SVAD task}. By employing our constructed CoT-AudioCaps data for \textbf{CoT-SFT, both strategies experienced further improvements in performance on the SVAD task}. 
Our goal is to align descriptions with the target audio, making CLAP the key metric. As shown in Table 2, CoT-SFT improves CLAP from 0.348 to 0.373 (Two-Stage) and 0.404 to 0.424 (Single-Stage). For Text-Text metrics, CoT-SFT shows a slight decline in BLEU\_1 and SPICE (word-level overlap and semantic errors), while outperforming in BLEU\_2, BLEU\_3, BLEU\_4, METEOR, ROUGE\_L, and CIDEr. This indicates that CoT-SFT improves matching of longer continuous phrases, enhances text similarity with synonyms and morphological variations, captures longer in-order subsequences even if they are non-contiguous, and aligns more closely with human consensus. 
As a result, the generated descriptions are more diverse, context-aware, and better aligned with human expectations for the target audio.
Furthermore, compared to the baselines AVCap-V and $\text{DALI}_{\text{OT}}^{\text{Att}}$, all metrics show significant improvements. These experimental results fully demonstrate the effectiveness of our methods.

\subsection{VT2A inference (RQ3)}
We utilized audio descriptions generated by our best-performing method VL2-CoT-SFT as text prompts for VT2A inference. The experimental results presented in Table \ref{tab:results2} indicate that STA-V2A\cite{ren2024sta} is more dependent on text prompts compared to FoleyCraft\cite{zhang2024foleycrafter}, thus experiencing a more significant performance decline when text prompts are absent. 
When utilizing audio descriptions generated by pre-trained VLMs as text prompts, although they perform better than having no text prompts at all, they still significantly underperform compared to the use of ground truth text prompts.
However, when using audio descriptions reasoned from silent videos through VL2-CoT-SFT, the generated audio showed improvements across all metrics compared to those generated with audio descriptions from pre-trained VLMs, \textbf{effectively narrowing the performance gap} with those using GT as prompts. The experimental results demonstrate that the \textbf{VL2-CoT-SFT method is an effective solution for addressing the challenge of lacking audio descriptions during VT2A inference}.

\section{Conclusion}
\label{sec:conclusion}
This paper introduces a new SVAD task that reasons audio descriptions from silent videos, tackling the challenge of audio descriptions missing in VT2A inference. Through evaluation of the SVAD task, we reveal VLMs' inherent limitations in modal-mismatch reasoning when target modalities are absent, and propose an innovative CoT-SFT strategy with our constructed CoT-AudioCaps dataset. Comprehensive experiments demonstrate that our CoT-SFT approach significantly enhances VLMs' reasoning capabilities in SVAD and the proposed method successfully addresses the challenge during VT2A inference. Future work will explore more techniques like process reward models to enhance VLMs reasoning capabilities in SVAD.

\newpage

\bibliographystyle{IEEEtran}
\bibliography{mybib}

\end{document}